\documentstyle{mn}
\newif\ifAMStwofonts

\def\msun{M$_{\odot}$ }
\def\lsun{L$_{\odot}$}

\input epsf

\ifoldfss
  \ifCUPmtlplainloaded \else
    \NewTextAlphabet{textbfit} {cmbxti10} {}
    \NewTextAlphabet{textbfss} {cmssbx10} {}
    \NewMathAlphabet{mathbfit} {cmbxti10} {} 
    \NewMathAlphabet{mathbfss} {cmssbx10} {} 
  \fi
  \ifAMStwofonts
    \ifCUPmtlplainloaded \else
      \NewSymbolFont{upmath} {eurm10}
      \NewSymbolFont{AMSa} {msam10}
      \NewMathSymbol{\upi}     {0}{upmath}{19}
      \NewMathSymbol{\umu}     {0}{upmath}{16}
      \NewMathSymbol{\upartial}{0}{upmath}{40}
      \NewMathSymbol{\leqslant}{3}{AMSa}{36}
      \NewMathSymbol{\geqslant}{3}{AMSa}{3E}

      \let\geq=\geqslant 
    \fi
  \fi
\fi 

\ifnfssone
  \newmathalphabet{\mathit}
  \addtoversion{normal}{\mathit}{cmr}{m}{it}
  \addtoversion{bold}{\mathit}{cmr}{bx}{it}
  \newmathalphabet{\mathbfit} 
  \addtoversion{normal}{\mathbfit}{cmr}{bx}{it}
  \addtoversion{bold}{\mathbfit}{cmr}{bx}{it}
  \newmathalphabet{\mathbfss} 
  \addtoversion{normal}{\mathbfss}{cmss}{bx}{n}
  \addtoversion{bold}{\mathbfss}{cmss}{bx}{n}
  \ifAMStwofonts
    \ifCUPmtlplainloaded \else
      \UseAMStwoboldmath
      \makeatletter
      \new@mathgroup\upmath@group
      \define@mathgroup\mv@normal\upmath@group{eur}{m}{n}
      \define@mathgroup\mv@bold\upmath@group{eur}{b}{n}
      \edef\UPM{\hexnumber\upmath@group}
      \new@mathgroup\amsa@group
      \define@mathgroup\mv@normal\amsa@group{msa}{m}{n}
      \define@mathgroup\mv@bold\amsa@group{msa}{m}{n}
      \edef\AMSa{\hexnumber\amsa@group}
      \makeatother
      \mathchardef\upi="0\UPM19
      \mathchardef\umu="0\UPM16
      \mathchardef\upartial="0\UPM40
      \mathchardef\leqslant="3\AMSa36
      \mathchardef\geqslant="3\AMSa3E

      \let\geq=\geqslant 
    \fi
  \fi
\fi 

\ifnfsstwo
  \DeclareMathAlphabet{\mathbfit}{OT1}{cmr}{bx}{it}
  \SetMathAlphabet\mathbfit{bold}{OT1}{cmr}{bx}{it}
  \DeclareMathAlphabet{\mathbfss}{OT1}{cmss}{bx}{n}
  \SetMathAlphabet\mathbfss{bold}{OT1}{cmss}{bx}{n}
  \ifAMStwofonts
    \ifCUPmtlplainloaded \else
      \DeclareSymbolFont{UPM}{U}{eur}{m}{n}
      \SetSymbolFont{UPM}{bold}{U}{eur}{b}{n}
      \DeclareSymbolFont{AMSa}{U}{msa}{m}{n}
      \DeclareMathSymbol{\upi}{0}{UPM}{"19}
      \DeclareMathSymbol{\umu}{0}{UPM}{"16}
      \DeclareMathSymbol{\upartial}{0}{UPM}{"40}
      \DeclareMathSymbol{\leqslant}{3}{AMSa}{"36}
      \DeclareMathSymbol{\geqslant}{3}{AMSa}{"3E}

      \let\geq=\geqslant 
    \fi
  \fi
\fi 

\ifCUPmtlplainloaded \else
  \ifAMStwofonts \else 
    \def\upi{\pi}
    \def\umu{\mu}
    \def\upartial{\partial}
  \fi
\fi

\title{Stellar Populations and Ages of M82 Super Star Clusters}
\author[J. S. Gallagher and L. J. Smith]
       {John S. Gallagher, III$^1$ and Linda J. Smith$^2$ \\
     $^1$Department of Astronomy, University of Wisconsin-Madison,
        5534 Sterling, 475 North Charter St.,\\ 
        Madison WI 53706, USA; jsg@astro.wisc.edu\\
     $^2$Department of Physics and Astronomy, University College
London, Gower Street, London, WC1E 6BT; ljs@star.ucl.ac.uk}
\date{Accepted 1998 December 8.
      Received 1998 November 23;
      in original form 1998 August 27}

\pagerange{\pageref{firstpage}--\pageref{lastpage}}
\pubyear{1998}

\begin{document}

\maketitle

\label{firstpage}

\begin{abstract}
We present high signal-to-noise optical spectra of two luminous super
star clusters in the starburst galaxy M82. The data for cluster F and
the nearby, highly reddened cluster L were obtained with the  William
Herschel Telescope (WHT) at a resolution of 1.6\,\AA. The blue
spectrum (3250--5540\,\AA) of cluster F  shows features typical of
mid-B stars. The red spectra (5730--8790\,\AA) of clusters F and L
show the Ca\,II triplet and numerous F and G-type absorption features. 
Strong Ca\,II and Na\,I interstellar absorption lines arising in M82
are also detected, and the $\lambda6283$ diffuse interstellar band
appears to be present. The quality of the WHT spectra allows us to
considerably improve previous age estimates for cluster F. By
comparing the blue spectrum with theoretical model cluster spectra
using the PEGASE spectral synthesis code (Fioc \& Rocca-Volmerange
1997), we derive an age of $60\pm20$ Myr. The strength of the Ca\,II
triplet is also in accord with this age. Cluster L
appears to have a similar age, although this is much less certain. The
measured radial velocities for the two clusters differ substantially,
indicating that they are located
in different regions of the M82 disk. Cluster F appears to be
deep in M82, slightly beyond the main starburst region while the
highly obscured cluster L lies near the outer edges of the disk. We
derive an absolute magnitude $M_V = -16.5$ for F indicating that it is
an extremely massive cluster. The presence of such a luminous super
star cluster suggests that the M82 starburst experienced an episode of
intense star formation $\approx 60$~Myr ago.
\end{abstract}


\section{Introduction}

M82 is one of the nearest examples of a spectacular starburst
galaxy. At a distance of 3.6~Mpc (see Freedman et al. 1994) its far
infrared luminosity is about 3.2$\times$10$^{10}$\lsun (Colbert et al.
1998), and the many
effects of the starburst, including extreme optical surface brightness
and giant outflows, are immediately apparent (e.g., Lynds \& Sandage 1963).
M82 has naturally played a special role in studies of intense star
formation on large scales (e.g., Telesco 1988).  Among the results
from ground-based optical and infrared investigations of M82 was the
recognition that dense, luminous star clusters are common in M82
(O'Connell \& Mangano 1978; hereafter OM78).  High angular resolution
optical imaging with the Wide Field Planetary Camera on the {\it
Hubble Space Telescope} ({\it HST}) revealed more than 100 clusters
located in the optically visible skin of the M82 starburst region
(O'Connell et al. 1995).  Infrared imaging penetrates the dense gas
and dust in the starburst and finds populations of compact, luminous
star clusters in the actively star-forming regions of the galaxy
(Satyapal et al. 1995, 1997).  Star clusters are evidently a featured
product of the intense star formation now occurring within M82.  This
is consistent with the more extensive results from the {\it HST}
showing compact, luminous `super star clusters' (SSCs) to be
frequently associated with starbursts under a wide range of conditions
(e.g., Whitmore \& Schweizer 1993;
Whitmore et al. 1995; Conti \& Vacca 1994; O'Connell, Gallagher \&
Hunter 1994; Meurer et al. 1995; Schweizer et al. 1996; see Ho 1997 for a
review).

The presence of super star clusters in starburst galaxies provides us 
with a useful tool for exploring some aspects of at least their recent 
star formation histories, while also yielding some information about 
chemical abundances in stars.  These measurements are feasible because 
star clusters closely approximate a coeval, single metallicity simple 
stellar population (SSP).  An SSP is the simplest class of stellar 
system to model, and one in which age and metallicity can be extracted 
from the integrated spectrum (see Bruzual \& Charlot 1993, Bruzual 1996). 
In return for their increased complexity, SSCs offer several practical 
advantages as compared with observations of individual stars; most 
significantly they can be 10--100 or more times as luminous as the 
brightest supergiant stars, allowing SSCs to be observed at greater 
distances or against more complex backgrounds.

Achieving the potential of SSCs as astrophysical probes of starburst
galaxies, however, has proven difficult. Starbursts are often dusty,
limiting the utility of optical and ultraviolet colors to yield ages
of younger clusters (e.g., Watson et al.  1996). For intermediate age
or older star clusters the well-known age-metallicity degeneracy
becomes an issue (Searle, Wilkinson \& Bagnuolo 1980). Optical and
ultraviolet spectra are available for a few blue SSCs, but initial
studies were limited to qualitative discussions (e.g., Arp \& Sandage
1985, Lamb et al. 1985). More recent observations have led to
considerable progress in analyzing very young SSCs, where strengths
and profiles of ultraviolet atomic resonance lines are powerful
diagnostics (e.g. Vacca et al. 1995, Gonz\' ales Delgado et al. 1997).

Optically luminous SSCs tend to be dominated in the blue by B- or
A-type spectra (Schweizer \& Seitzer 1993, Zepf et al. 1995; Brodie et
al. 1998). As a result, ages are not readily determined from
measurements of individual spectra features, especially when the
Balmer jump is not available (Diaz 1988).  In addition, these studies
have found that the observed strengths of the Balmer lines exceed those
predicted by the Bruzual \& Charlot (1993) models. Brodie et al. (1998)
find that a truncated IMF is needed to reproduce their observations
of the galaxy NGC 1275. For the M82 starburst,
Satyapal et al. (1997) use a wealth of infra-red data to infer a
typical age of $10^7$~yr for the clusters within a radius of 250~pc of
the nucleus. They also find an age dispersion of $6 \times 10^6$~yr
which is correlated with the projected distance of the clusters from
the nucleus, and therefore suggest that the starburst is propagating
outwards from the nucleus.

In this paper we present a new spectroscopic study of the luminous SSC
F in M82 (OM78) and the nearby, highly reddened cluster L (Kronberg,
Pritchet \& van den Bergh 1972). Cluster F has a size of about
9$\times$5~pc from observations with the original Planetary Camera on
{\it HST} and an estimated luminosity of $M^0_V \approx-$14.5
(O'Connell et al. 1995), and is therefore one of the most luminous
younger star clusters known.  It is located 25 arc sec ($=440$~pc)
south-west of the 2.2 $\mu$m nucleus and is considered to be in a less
obscured region of the starburst (O'Connell et al. 1995).  The goal of
our program is to develop improved approaches for quantitatively
matching spectra of SSCs to model predictions as a means to determine
cluster ages. Observations for this project were made at the 4.2\,~m
William Herschel Telescope (WHT) using the double beam spectrograph
{\it ISIS}.  This allowed us to obtain wide wavelength coverage with
good spectral resolution, and thereby considerably improve the
estimated age for cluster F.

The observational procedures are reviewed in the next section. In Sect. 3 
we describe the spectra, while Sect. 4 compares observed spectra with 
predictions from a population synthesis model. Results are discussed 
in Sect. 5, including implications for the duration of the M82 starburst 
and the quality of current models for SSPs with near solar metallicity.

\section{Observations and Data Reduction}
\begin{table}
\centering
\begin{minipage}{80mm}
\caption {Journal of WHT Observations.}
\label{tab1}
\begin{tabular}{@{}cccc}
\hline
\multicolumn{2}{c}{Blue Arm} & \multicolumn{2}{c}{Red Arm}\\
Wavelength & Exposure & Wavelength & Exposure \\
Range & Time & Range & Time \\
(\AA) & (s) & (\AA) & (s) \\[10pt]
\hline 
3234--4042 & 2400 & 5726--6534 & 1200 \\
     &      & 6478--7286 & 1200 \\
3984--4792 & 1200 & 7224--8032 & 1200 \\
4729--5537 & 2400 & 7977--8785 & 2400 \\
\hline
\end{tabular}
\end{minipage}
\end{table}
\begin{figure}
\epsfxsize=230pt
\epsfbox[63 393 410 627]{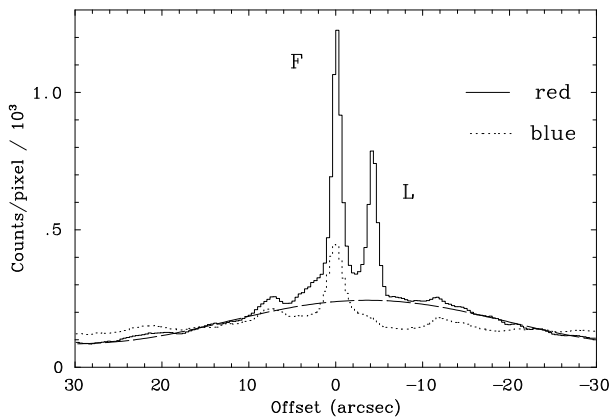}
\caption{The raw spatial profiles for the red wavelength region
(8100--8600\,\AA; solid line) and the blue region (4100--4600\,\AA;
dotted line) normalised to the same exposure time.  The two
superclusters F and L are indicated.  The x-axis represents the
spatial offset with respect to the centre of SSC F and SSC L is $4''$
to the SW of F.  The dashed line shows a representation of the
polynomial fit used to subtract off the underlying galaxy background
for the red spectrum.}
\label{fig1}
\end{figure}
Observations of SSC F in M82 were obtained on 25/26 March 1997 with
the WHT on La Palma, Canary
Islands. SSC F was observed at a position angle of $20^\circ$ so that
spectra of SSC L could also be obtained.  The journal of observations
is given in Table~\ref{tab1}.  The double-beam spectrograph ISIS was
used with Tektronix CCDs ($1024 \times 1024$ 24$\mu$m pixels) on both
the blue and red arms. Two 600 line gratings were used to give a
continuous wavelength coverage from 3234--5537\,\AA\ and
5726--8785\,\AA.  A long slit of dimensions $1''.0 \times 4'.0$ was
chosen to give a two-pixel spectral resolution of 1.6\,\AA. The
spatial scale was $0''.36$\,pixel$^{-1}$ and an average seeing of
$1''.2$ was measured over the period of the observations.

The data were reduced using the PAMELA optimal extraction routines
(Horne 1986) within the FIGARO (Shortridge et al. 1997) software package.
Each frame was first bias-corrected, and those obtained at the reddest
wavelength setting were divided by a normalised flat-field to remove
the fringing.  In Fig. \ref{fig1}, we show the spatial profile along
the slit at representative red and blue wavelengths obtained by
collapsing the raw images from 8100--8600\,\AA\ and
4100--4600\,\AA. It can be seen that SSC L is not detected in the blue
region, and that there is a substantial contribution from the galaxy
background. This was subtracted by using polynomial fits to each
pixel in the spatial direction, as shown in Fig. \ref{fig1}.  The
object data were then optimally extracted and wavelength calibrated
using Cu-Ar comparison spectra taken before or after each integration.
The red arm spectra were next corrected for telluric absorption
features using a B star standard, and all spectra were
extinction-corrected using the curve appropriate to La Palma.

Finally the data were flux calibrated using the flux standard Feige 34
(Oke 1990) and the individual wavelength ranges were merged. To
accomplish this, small adjustments of 5--15\% in flux level were
required.  The final spectra have a signal-to-noise ratio per
pixel in the continuum of 30--40 above the Balmer jump for SSC F and
longward of 7000\,\AA\ for SSC L.
\begin{figure*}
\epsfbox[40 103 489 683]{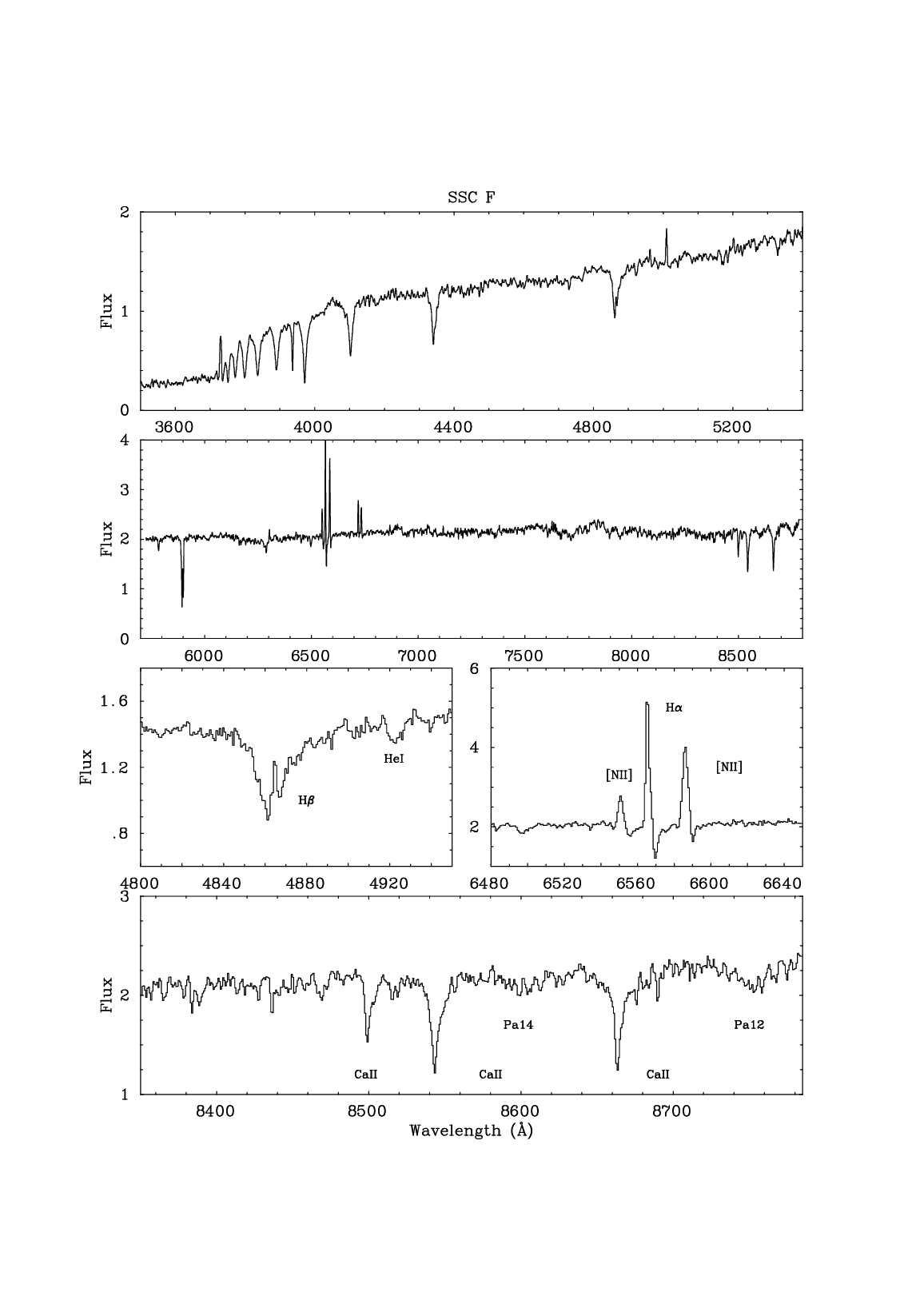}
\caption{The blue and red spectra of SSC F
(smoothed with $\sigma=1.0$\,\AA\ for presentation purposes).
 The lower three boxes (at the full spectral resolution) show
enlargements covering the H$\beta$, H$\alpha +$ [N\,II], and Ca\,II
triplet regions.  The y-axis is in units of $10^{-15}$
ergs\,s$^{-1}$\,cm$^{-2}$\,\AA$^{-1}$ }
\label{fig2}
\end{figure*}
\begin{figure*}
\epsfbox[40 239 487 678]{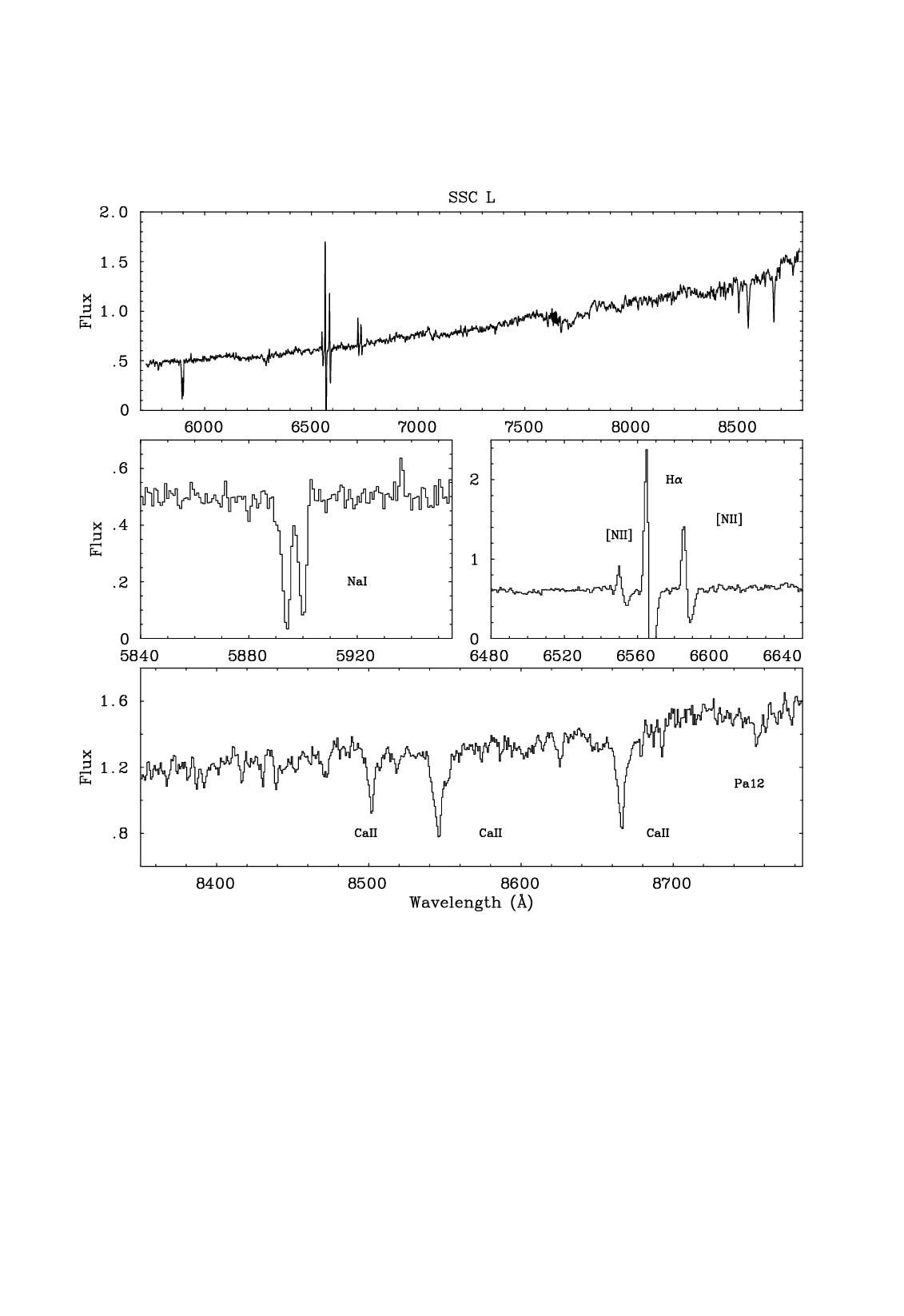}
\caption{The red spectrum of SSC L (smoothed with $\sigma=1.0$\,\AA\ for 
presentation purposes). The lower three boxes (at the full spectral
resolution) show
enlargements covering the Na\,I interstellar absorption lines,
H$\alpha +$ [N\,II], and Ca\,II triplet regions.  The y-axis is in
units of $10^{-15}$ ergs\,s$^{-1}$\,cm$^{-2}$\,\AA$^{-1}$ }
\label{fig3}
\end{figure*}
\begin{table*}
\centering
\begin{minipage}{160mm}
\caption {Velocity and Equivalent Width (W$_\lambda$) Measurements.}
\begin{tabular}{@{}lllrclrcl@{}}
\hline
Line & $\lambda_{\rm {lab}}$ & \multicolumn{3}{c}{SSC F} & 
\multicolumn{3}{c}{SSC L} & Comments \\
&& Velocity & Measured & Window$^1$ 
& Velocity & Measured & Window$^1$ \\
& (\AA) &(km\,s$^{-1}$) & W$_\lambda$ (\AA) & W$_\lambda$ (\AA) &
(km\,s$^{-1}$) & W$_\lambda$ (\AA) & W$_\lambda$ (\AA)\\[10pt]
\hline 
{[}O\,II{]} & 3727 & $+176$ & $-8.81\pm0.21$ & &&&& Nebular\\
H12     & 3750.1 & $+41$  & $3.03\pm0.13$\\
H11     & 3770.6 & $+72$  & $5.01\pm0.16$\\
H10     & 3797.9 & $+108$ & $6.59\pm0.17$ &6.55&\\
H9      & 3835.4 & $+56$  & $7.10\pm0.20$ &7.16&\\
H8      & 3889.1 & $+59$  & $6.73\pm0.15$ &7.00&\\
Ca\,II  & 3933.7 & $+191$ & $2.43\pm0.09$ &2.45& &&&IS\\
H$\epsilon$ & 3970.1 & $+70$ & $8.67\pm0.16$ & 8.56& \\
H$\delta$ & 4101.8 & $+40$ & $8.17\pm0.24$ & 7.97& \\
H$\gamma$ & 4340.5 & & $6.28\pm0.17$ & 6.02& \\
H$\gamma$ & 4340.5 & $+187$ &&&&&&Nebular \\
He\,I+Fe\,I,II & 4385 & & $0.43\pm0.07$ & 0.14 \\
He\,I     & 4471.5 & $+57$ & $0.40\pm0.07$ & 
\raisebox{-1.0ex}{1.09} \\
Mg\,II    & 4481.1 & $+77$ & $0.34\pm0.08$ \\
H$\beta$  & 4861.3 &  & $6.55\pm0.14$ & 6.08& \\
H$\beta$  & 4861.3 & $+199$ &&& &&&Nebular \\
He\,I     & 4921.9 & $+27$ & $0.56\pm0.08$ & 0.35 \\
{[}O\,III{]} & 4958.9 & $+181$ & $-0.36\pm0.06$ & &&&& Nebular \\
{[}O\,III{]} & 5006.8 & $+164$ & $-0.85\pm0.05$ & &&&& Nebular \\
Mg\,I     & 5170   && $0.76\pm0.12$ & \raisebox{-1.0ex}{1.20$^2$} \\
Mg\,I     & 5185.5 &$+23$ & $0.43\pm0.10$ \\
Fe\,I$+$Ca\,I& 5270&& $0.40\pm0.04$ & 0.57$^2$\\
Cr\,I$+$Cu\,I & 5782 & & $0.75\pm0.08$ & 0.32 &
                       & $1.09\pm0.17$ & 2.53\\
Na\,I   & 5890.0 & $+193$ & \raisebox{-1.0ex}{$6.20\pm0.10$} & 
\raisebox{-1.0ex}{6.49} & $+200$ & 
\raisebox{-1.0ex}{$6.59\pm0.18$} & \raisebox{-1.0ex}{6.74} & IS\\
Na\,I   & 5895.9 & $+183$ & & & $+198$ & & &IS\\
&         6283.9 & $+206$ & $1.52\pm0.11$ & & $+174$ & $2.17\pm0.22$ & & DIB?\\
{[}O\,I{]} & 6300.3 & $+181$ & $-0.48\pm0.06$ &&  & &&Nebular \\
Fe\,I$+$Ba\,II$+$Ca\,I & 6497 & & $0.79\pm0.07$ & 0.52 &&
$0.82\pm0.16$ & 0.71 \\
{[}N\,II{]} & 6548.0 & $+111$ & $-1.34\pm0.05$ && 
$+80$& $-0.86\pm0.08$ &&Nebular \\
H$\alpha$ & 6562.8 & $+104$ & $-4.34\pm0.05$ && 
$+77$ & $-6.05\pm0.08$ &&Nebular  \\
{[}N\,II{]} & 6583.4 & $+108$ & $-3.59\pm0.05$ && 
$+61$ &$-2.98\pm0.09$ &&Nebular \\
{[}S\,II{]} & 6716.5& $+126$ & $-1.34\pm0.06$ && 
$+65$ & $-1.57\pm0.09$ &&Nebular  \\
{[}S\,II{]} & 6730.9 & $+121$ & $-1.10\pm0.06$ && 
$+61$& $-1.14\pm0.09$ && Nebular \\
Ti\,I$+$Fe\,I & 8468 && $0.54\pm0.08$ & & & $0.60\pm0.08$\\
Ca\,II  & 8498.0 & $+31$ &$1.43\pm0.13^3$&& $+133$ & $1.21\pm0.14^3$\\
Fe\,I$+$Ti\,I & 8516 & & $0.41\pm0.08$ & & & $0.83\pm0.09$\\
Ca\,II  & 8542.1 & $+36$ &$4.09\pm0.13^3$ && $+126$ & $4.14\pm0.14^3$\\
Pa 14   & 8598.4 && $1.51\pm0.16$ \\
Ca\,II  & 8662.1 & $+38$ &$4.41\pm0.13^3$ && $+135$ & $4.75\pm0.14^3$\\
Fe\,I   & 8688.6 & $+38$ &$0.36\pm0.06$ & & $+137$ & $0.51\pm0.07$\\
Pa 12   & 8750.5 &       &$2.56\pm0.16$ & 2.45 & & $2.16\pm0.18$ &
2.30 \\
\hline
\noalign{$^1$ all windows as defined by Bica \& Alloin (1986)
except for the following: $^2$ windows as defined by Faber et al.
(1985); $^3$equivalent widths measured using the bands defined 
by Diaz et al. (1989).}
\end{tabular}
\label{tab2}
\end{minipage}
\end{table*}
\section{Description of Spectra}
\subsection{Continuum}
In Figs. 2 and 3 we show the flux calibrated spectra for SSC's F and L
and enlargements over selected wavelength ranges to highlight various
spectral features.
We have performed
synthetic photometry on the spectra to derive magnitudes and colours.
For SSC F, we derive $UBV$ magnitudes of 17.6, 16.9 and 15.8
respectively giving $(U-B)=0.70$ and $(B-V)=1.07$. These colours are
in excellent agreement with those of OM78 who
derive $V=15.00$, $(U-B)=0.69$ and $(B-V)=1.04$, using a $7''.5$
aperture.  From HST images, O'Connell et al. (1995) find $V=16.3$
within a radius of $0''.5$.  For SSC L, we derive a monochromatic R
magnitude of 16.7 compared to 15.6 for F.

\subsection{Absorption Lines}
Line identifications, heliocentric velocities and equivalent widths
of the principle spectral features are given in Table 2 for SSC's F
and L. The equivalent widths were measured by defining continuum
windows either side of the features and the errors given in Table 2
represent $1\sigma$ uncertainties based on the signal-to-noise in
the continuum windows. We also list equivalent widths based on
the windows defined by Bica \& Alloin (1986) which tend to be much
wider than the actual spectral features present in our data.
We now discuss each SSC in turn.

\subsubsection{M82 SSC F}
The blue spectrum is dominated by broad Balmer absorption lines and a
strong Balmer jump. Comparison of the equivalent widths of H9, H8
and H$\delta$ (least likely to be affected by nebular emission) with 
main sequence stars in the Jacoby, Hunter \& Christian (1984) spectral 
library shows that the best match is for a  mid-B spectral type. 
This classification is
confirmed by the detection of weak He\,I lines (e.g.
$\lambda4922$ to the red of H$\beta$ in Fig. 2.) and Mg\,II
$\lambda4481$. We find that the strong and narrow Ca\,II K line arises
from  interstellar gas in M82 rather than stellar atmospheres since
its velocity agrees with that of the Na\,I D lines and is very different to
the stellar absorption features.

In the red spectrum, the strongest absorption features present belong
to the Ca\,II triplet and we also detect broad absorption from members
of the Paschen series. The equivalent widths given in Table 2 for the
Ca\,II triplet have been measured using the continuum and line windows
of Diaz, Terlevich \& Terlevich (1989).  The combined strength of the
two strongest members (Ca2$+$Ca3) is 8.5\,\AA, uncorrected for Paschen
line absorption.  The mean velocity of the triplet is
$+$35\,km\,s$^{-1}$ and probably represents the most accurate value
for the radial velocity of SSC F since the Balmer lines are broad and
contaminated by nebular emission.

The overall spectrum redward of 5000\,\AA\ has various
features attributable to F and G stars (Andrillat, Jaschek \& Jaschek 1995;
Carquillat et al. 1997). In addition to the Ca\,II triplet,
we detect weak absorption features due to Cr\,I$+$Cu\,I $\lambda5782$,
Fe\,I$+$Ba\,II$+$Ca\,I $\lambda6497$ as well as various Ti\,I and Fe\,I lines
near the Ca\,II triplet. No molecular bands 
due to e.g. TiO are detected in our spectra.

\subsubsection{M82 SSC L}
The red spectrum (Fig. 3) clearly shows that the Ca\,II triplet is
present and a weak Pa 12 absorption line is detected, suggesting that
blue stars are present in this cluster.  Weak low ionization
absorption features arising from F and G stars are also detected and
have similar strengths to those found in F.  The strengths of the
Ca\,II triplet lines are similar to those measured for SSC L with
Ca2$+$Ca3 $=8.9$\,\AA. The radial velocities are, however, quite
different with a mean for the Ca\,II triplet of $+$131\,km\,s$^{-1}$
compared to $+$35\,km\,s$^{-1}$ for F.  This large difference in
radial velocity indicates that clusters L and F lie at different
locations along our sight line through the disk of M82.

\subsection{Emission Lines}
For SSC F, we detect nebular emission lines of H$\alpha$, H$\beta$,
H$\gamma$, [O\,II] $\lambda3727, 3729$, [O\,III] $\lambda4959, 5007$,
[O\,I] $\lambda6300$, [N\,II] $\lambda6548,6584$, and [S\,II]
$\lambda6717,6731$. The H$\beta$ and H$\alpha$ $+$ [N\,II] emission
line profiles are shown in Fig. 2 and heliocentric radial velocities
and equivalent widths are given in Table 2. For the H$\beta$,
H$\gamma$, [O\,I], [O\,II], [O\,III] lines, we derive a mean velocity
of $+181\pm12$\,km\,s$^{-1}$ which agrees well with the mean velocity
of the Ca\,II K and Na\,I D interstellar lines of
$+189\pm5$\,km\,s$^{-1}$. The H$\alpha$, [N\,II] and [S\,II] lines show a
lower velocity of $+114\pm9$\, km\,s$^{-1}$ in good agreement with
that determined by  OM78 of
$+121\pm10$\,km\,s$^{-1}$ for the same features. Inspection of the
images shows that the spatial structure of these lines is more
complicated than the [O\,II] or [O\,III] emission which is constant in
intensity and velocity across the region displayed in Fig 1. The
emission for the strong H$\alpha$, [N\,II] and [S\,II] lines is clumpy
with two velocity components -- the component seen in the other lines
at $+181$\,km\,s$^{-1}$  and another component at $+114$\,km\,s$^{-1}$
which lies between SSCs F and L. As a result of this structure, the
strong $+181$\,km\,s$^{-1}$ component is over-subtracted from both SSC
spectra and appears in absorption in Figs. 2 and 3.

For SSC L, we detect nebular emission lines of H$\alpha$, [N\,II] and
[S\,II] and find compared to SSC F a lower mean heliocentric velocity
of $+69\pm9$\,km\,s$^{-1}$. The equivalent widths of the nebular lines
are similar for the two spectra.

\subsection{Diffuse Interstellar Bands}
Due to their moderate-to-high levels of interstellar dust obscuration,
M82 clusters F and L provide excellent sight lines for measuring
interstellar features in the spectrum of M82. We have already noted
the presence of strong interstellar absorption from the atomic species
in the Ca\,II K and Na\,I D lines.  Our spectra also have sufficient
resolution and signal-to-noise ratios to allow sensitive searches for
diffuse interstellar bands (DIBs). Major features of DIBs are reviewed
by Herbig (1995). The strongest DIBs in the optical spectral region
are the $\lambda$5778-80, $\lambda$6283, and $\lambda$6533 features,
and we therefore looked for these in our spectra. The
$\lambda$5778--80 feature is coincident with the $\lambda5782$
Cr\,I$+$Cu\,I blend which we positively identify in the spectra of F
and L. The $\lambda$6283 DIB, is, however, probably present. We
see a weak feature at 6288\,\AA\ (see Table 2) which is definitely
real as it is occurs in the overlap region between spectra and is seen
in both sets of data.  Comparison with the spectral library of Jacoby
et al. (1984) shows that it does not appear to be a stellar absorption
feature. Likewise, comparison with the sky spectrum and the atmospheric
standard shows that it is not due to over-subtracted sky or a telluric
feature.
The strength of the $\lambda$6288 absorption also
correlates with reddening, being stronger in cluster L, as expected
if this is a DIB.
The possibility therefore remains that the feature at $\lambda6288$
could be a DIB arising from the interstellar medium of M82. 

The absorption strengths of DIBs are complicated. They are usually 
strongest in the boundaries of molecular clouds and tend to weaken 
in purely molecular regions as well as being absent from H\,II regions 
(e.g. Jenniskens et al. 1994).  This behavior is consistent with models 
where DIBs are produced by ionized large molecules (Sonnentrucker et al. 
1997). Typical Galactic paths with E(B$-$V)$\approx$1 produce central 
absorption depths of $\approx 0.75$ in the $\lambda6283$ DIB
(e.g. Jenniskens \& Desert 1993, Herbig 1995); slightly stronger than 
what we observe in either cluster F (E(B$-$V)$\approx$1--1.5) or 
cluster L (E(B$-$V)$\geq$2) in M82. 
The modest strengths of the DIBs along paths 
through the M82 ISM which contains strong Na I D lines 
and thus neutral gas suggests that we 
are observing these two clusters through regions 
with few ionization boundaries. In this case we might 
expect small scale structure to be common in the ISM, which could 
contribute to non-uniform interstellar obscuration 
across the faces of the clusters. 
Follow-up observations are needed to establish whether
M82 is the second galaxy outside of the Local Group where DIBs have
been detected (the first being supernova 1986G in NGC~5128; see
Herbig 1995).

\begin{figure*}
\epsfbox[67 253 484 538]{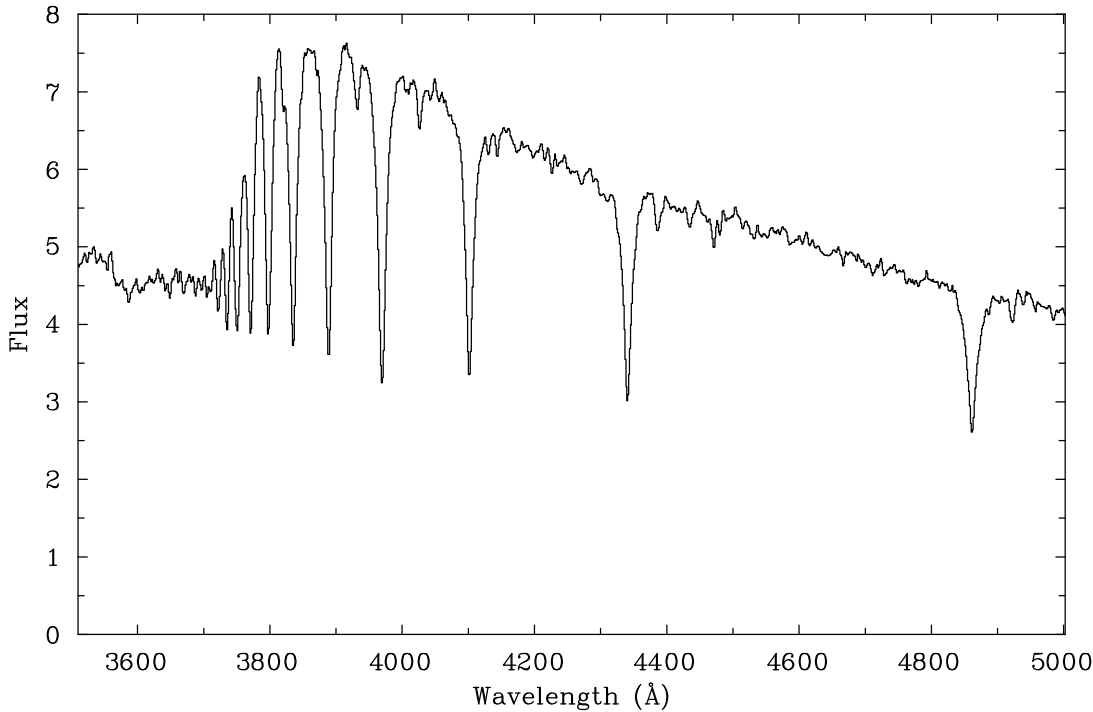}
\caption{This plot shows the model spectrum for a star cluster with 
an age of 60~Myr computed with PEGASE. The flux is in units of 
erg~s$^{-1}$~\AA$^{-1}$ per solar mass of cluster where  
a Salpeter initial mass function was assumed. The spectrum was synthesized 
using the Jacoby et al. (1984) library of stellar spectra 
and Geneva stellar evolution tracks}
\label{60M_model}
\end{figure*}
\begin{figure*}
\epsfbox[67 253 484 538]{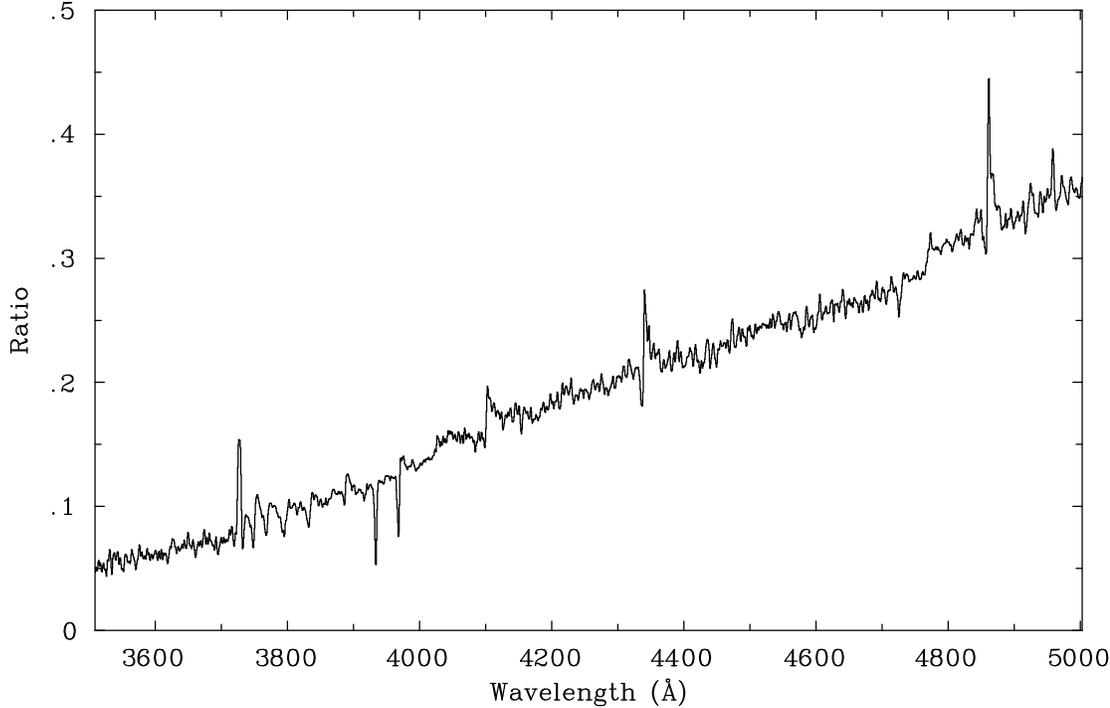}
\caption{The ratio is plotted between the observed cluster F blue
spectrum and 60~Myr PEGASE model displayed in Fig.~\ref{60M_model}.  The
ratio gives a model-dependent estimate for the mass of the cluster
required to provide the observed flux, and in the absence of dust
obscuration should be a constant function of wavelength. This ratio is
normalized such that a ratio of 1 would correspond to a model cluster
mass of 1.6$\times$10$^6$ \msun}
\label{MR}
\end{figure*}
\section{Comparisons with Theoretical Star Cluster Models}
\subsection{Blue wavelength region}
Quantitative models now exist for spectra of single-age, solar metallicity 
star clusters which formed over short time scales. We used the results from 
the spectral synthesis code  `Projet d'Etude des Galaxies par 
Synth\`ese Evolutive' (PEGASE; Fioc \& Rocca-Volmerange 1997). These 
models are well-suited to our needs because they include supergiant 
stars that are critical in the spectra of young star clusters; models 
that exclude such stars will not give proper matches to integrated 
spectra of younger super star clusters (e.g. Brodie et al. 1998). 
We selected the `Geneva' stellar evolution tracks from 
Schaller et al. (1992) and Charbonnel et al. (1996) in computing 
cluster models with PEGASE.

PEGASE also offers a choice of spectra synthesized from relatively low
spectral resolution model atmospheres, or the Jacoby et al.  (1984)
spectrophotometric library. We chose the latter option for this
project, as the near solar metallicity should be a reasonable match to
young objects in M82, and we desired the best possible spectral
resolution.  Even though this appears to still be the highest spectral
resolution Galactic spectrophotometric library that includes a wide
range of spectral and luminosity classes, it has lower resolution
(4.5\AA), and often lower per pixel signal-to-noise ratios, than our
WHT spectra of M82 cluster F.  Improved spectral libraries will be
important for future comparisons between model and observed spectra of
nearby SSCs.

Before comparing with blue-region spectra of model star clusters, 
we  smoothed the cluster F spectrum with a Gaussian having
$\sigma=1.0$~pixel
over $\lambda \lambda$3510--5002\,\AA.  
This slightly degrades our resolution to about 2 \AA.
We also shifted the observed spectrum by $-175$~km~s$^{-1}$  to account 
for the approximate redshift of the stellar absorption lines. The 
IRAF package `sarith' was then used to make ratios between the 
wavelength- and flux-calibrated data and lower 
spectral resolution blue region PEGASE models.
Models were computed for ages of 20, 40, 60, 80, and 100~Myr with a 
Salpeter IMF. The best-fitting 60~Myr model blue region 
spectrum is shown in Fig.~\ref{60M_model}.

The comparisons between PEGASE models and data are very good. Ratios 
between observations and model 
spectra reveal both components of the interstellar H and K Ca\,II absorption 
lines, as well as fainter Balmer emission from  the H$\gamma$ and H$\delta$ 
lines. Some `ringing' is present among the higher Balmer series; 
probably due to the modest spectral resolution of the spectral 
library as well as mismatches between wavelength scales.  

The ratios of  spectra display an upward slope towards the red caused by 
dust obscuration in M82. OM78 estimated the reddening 
towards cluster F to be E(B$-$V)$\approx$1; our model fits suggest 
a slightly higher value of E(B$-$V)$\approx$1.5. However, a standard 
Galactic extinction model does not properly correct the data; this 
law has too much curvature in the near UV; there is likely to be dust near 
cluster F and possibly even patchy obscuration (cf. 
conditions in the starburst nucleus of NGC~253; Watson et al. 
1996), so it is not surprising that the Galactic extinction model fails 
(see also Calzetti 1998). Because of this uncertainty in the proper choice 
of an obscuration correction, we cannot use the continuum spectral energy 
distribution at optical wavelengths to determine the cluster's age.

Our efforts therefore concentrated on matching the Balmer absorption 
lines and gravity/temperature-sensitive Balmer jump as the basis for 
selecting best-fitting models. While the models do not perfectly 
reproduce our 
observations, they do not display gross difficulties of the type 
encountered by Brodie et al. (1998) in their efforts to fit blue 
spectra of SSCs in NGC~1275. 
We therefore had no need to invoke an unusual form of the initial mass 
function; this cluster is fitted adequately with a Salpeter IMF. The 40
Myr PEGASE model gives the best fit to 
the H$\beta$, H$\gamma$, and H$\delta$ absorption lines but does not 
do as well for the Balmer jump region as the 60 and 80~Myr models. 
The 20 and 100~Myr cluster models gave worse fits in all areas. We 
therefore adopt an age for cluster F of 60$\pm$20~Myr from the PEGASE model 
fitting.

The ratio between the observed cluster F spectrum and 60~Myr model is shown  
in Fig.~\ref{MR}. Since we are comparing a monochromatic luminosity with 
a predicted luminosity per unit cluster mass, the ratio is in units 
of solar masses. The normalization leads to a ratio of 1 in the plot 
corresponding to a cluster mass of $1.6\times 10^6$~\msun. 
If we correct by a factor of 40--200 for A$_B =$4--6 magnitudes, 
then we see that a model cluster mass of 
$>$10$^7$~\msun is needed to reproduce the blue luminosity. 
Because the form of the initial mass function in SSCs is 
poorly known, this mass is highly uncertain. However, it does indicate 
that cluster F is likely to be very massive, in agreement with 
results from earlier studies (OM78, O'Connell et al. 1995).

\subsection{Red wavelength region}
The prime diagnostic for measuring ages in the red part of the spectrum
is the strength of the Ca\,II triplet (CaT).  The presence of a strong
CaT is often used as an indicator of red supergiants (RSG) and therefore
an age of $10^7$~yr. The CaT is, however, strong in yellow
supergiants as well,
and as discussed by Mayya (1997), is {\it not\/} an unambiguous
signature of RSGs. Mayya (1997) finds that, at solar metallicity, the
evolution of the strength of the two strongest members of the 
CaT (Ca2$+$Ca3) shows two peaks -- one from RSGs
at $\sim 10^7$~yr ($W_\lambda \approx 11$\,\AA), and one at 60\,Myr
due to red giants and asymptotic branch stars ($W_\lambda \approx 7$\,\AA).

In M82 clusters F and L, the combined equivalent widths of Ca2$+$Ca3
are 8.5 and 8.9\,\AA\ respectively, uncorrected for Paschen line
absorption. Spectral synthesis models (Leitherer, priv. comm.) at
ages of 40--80\,Myr predict equivalent widths of 7.9--8.3\,\AA\
(including the Paschen lines) and 5.6--5.9\,\AA\ (pure CaT). The
strength of the CaT in the spectrum of SSC F is therefore in accord with
the age deduced from the blue spectrum of $\approx 60$\,Myr. For
SSC L, the only potential age information that we have is the strength
of the CaT. The similarity in the strength of this feature, and the
overall spectral appearance, indicates that SSC L is probably of a
similar age to F. 

\section{Discussion}

The combination of high quality optical spectra and good spectral synthesis 
models for star clusters has allowed us to derive an age for M82 SSC F. 
Since SSCs are hallmarks of episodes of intense star formation (see Hunter, 
O'Connell, \& Gallagher 1994; O'Connell et al. 1994;
Watson et al. 1996), cluster F provides one benchmark in the history of 
the M82 starburst. While this is only one star cluster, it is a massive 
object, and so its birth would likely have been noticeable on a 
galactic scale; the SFR for making this cluster during a $\approx$ 1~Myr 
dynamical time scale would have been $>$1~\msun yr$^{-1}$. Thus our 
data suggest a {\it minimum} age for the starburst in M82 of 40~Myr 
and a most probable age of $\geq$ 60~Myr.

Ages of the M82 starburst have been previously derived from integrated spectra 
of the central regions of this galaxy by Rieke et al. (1993). Their 
preferred models consist of two bursts within the past $\sim$30~Myr, 
each with a duration of about 5~Myr. Doane \& Mathews (1993) based their 
models on the energetics of the stellar populations, including supernova 
rates estimated from radio observations, and found the starburst has been 
active for 30--60~Myr. Satyapal et al. (1997) observed twelve luminous 
concentrations within the central starburst in the infrared. They identify 
these objects with embedded SSCs, whose frequent presence is also 
required to fit the numbers of radio supernovae remnants (Golla, 
Allen, \& Kronberg 1996). The ages of these SSCs with high near-IR 
luminosities are estimated by Satyapal et al. (1997) to be near 10~Myr, 
and they present a model in which the starburst has propagated outwards 
from the nucleus during this time period.

At larger radii there is evidence that the starburst is older. For
example, Shen \& Lo (1995) measured a change in the properties of 
the molecular medium between the central and mid-disk zones of M82. 
The molecular gas at a radius of 390~pc is much more disturbed than that 
in the inner 125~pc region, suggesting that the outer gas was disrupted 
by an earlier starburst. They identify this with the
30~Myr starburst event suggested by Rieke et al. (1993). OM78 also 
noted that the outer region is probably an older event than that 
in the centre.

The structure of M82 is an important factor when discussing the
``age'' of the starburst (see Telesco 1988). Cluster F lies to the
southwest of the M82 nucleus, slightly above its very heavily obscured
mid-plane (Larkin et al. 1994).  The rapid variation in extinction
across this area is also illustrated by the presence of the
highly-reddened SSC L near cluster F.  It is in the region of M82
where the near-IR surface brightness remains roughly constant, and may
be associated with a ring structure that is inferred to surround a
central bar (see Telesco et al. 1991, Larkin et al. 1994, Achtermann
\& Lacey 1995).  

We can also use the measured radial velocities of the absorption and
emission lines in the spectra of clusters F and L to obtain
information on their locations within M82. G\" otz et al. (1990) and
McKeith et al. (1993) present detailed studies of the structure of the
disk of M82 using optical and near-IR emission and absorption line
velocities. In common with previous studies, they find that the size
of the rotational velocity gradient increases with wavelength because
of the obscuration in M82. The heliocentric velocities we derive for
the interstellar absorption lines ($+189\pm5$\,km\,s$^{-1}$) and the
blue nebular emission lines ($+181\pm12$\,km\,s$^{-1}$; Sect. 3.3)
agree well with those determined for the same lines near the position
of cluster F by G\" otz et al. (1990). 
These velocities are below
the systemic value, but well above the rotation velocity in this region
of M82 (e.g. Weliachew et al. 1984, Achterman \& Lacey 1995),
consistent with a model where the emission lines originate in ionized
gas located on the near side of the outer disk of M82 (McKeith et al.
1993).

In contrast, the radial velocity of $+35$ km~s$^{-1}$ we derive for
cluster F is very near the expected rotation curve peak at its
position. Cluster F therefore should be located near the mid-point
of our sight line through the disk of M82, at a radial distance of about 
440~pc from the nucleus, near the edge of the
dense gaseous ring. Possibly a hole in the dust layer is allowing us to
see cluster F relatively deep within M82 (see Larkin et al. 1994). For
cluster L, we find a CaT velocity of $+131$\,km\,s$^{-1}$ which is
similar to our red H\,II region emission line velocities and those of
G\" otz et al. (1990). This suggests that L is located towards the
outer edges of the disk at a radius of somewhat more than 440~pc 
in a region of high obscuration.

The different ages derived for the M82 starburst then may simply reflect 
variations of the star formation history within this extraordinary galaxy.
If we assume that cluster F is a typical component of the `ring' or 
mid-disk, then the starburst occurred in this region rather long ago, 
and in the last 20-30~Myr has become more concentrated in the central 
part of the galaxy, where a strong bar may be present. 
This model is consistent with 
previous results in suggesting the earlier M82 starburst occurred in the 
mid-disk, and the more recent event is more centrally concentrated. 
Age dating of a larger population of star clusters beyond the 
currently active central-region starburst has the potential to provide 
a clearer view of the development of the M82 starburst.

\section{Conclusions}
We have obtained high signal-to-noise ratio optical spectra with
good wavelength resolution for star clusters F and L in M82.
Quantitative fits to theoretical models are successful in reproducing
the spectra with an SSP having an age of $\approx 60$\,Myr for cluster F.
This corresponds to a main sequence turnoff mass of about 7\,$M_\odot$,
consistent with the temperatures of stars associated with weaker
spectral lines seen in the blue region.  Cluster L appears to have a
similar age, but the result is much less certain due to the lack of
blue spectra.  The combination of high quality spectra and stellar
population models which include supergiant stars provides a powerful
means to analyze young super star clusters.

Combining our estimate of V=15.8 with an extinction
of E(B$-$V)$\approx1.5$ we derive $M_V =-16.5$ for cluster F, two
magnitudes brighter than that determined by O'Connell et al. (1995).
It therefore appears likely that these clusters are extremely massive
and may have the ability to survive as bound systems over long
time scales (e.g. Goodwin 1997). A measurement of the stellar velocity
dispersion in cluster F would be extremely useful in constraining its mass
and thus similarity to globular star clusters as well as
future evolution (cf. Ho \& Filippenko 1996).

The spectra indicate that while cluster L is more obscured than
cluster F, the latter star cluster likely lies near the outer edge
of the main starburst region in M82.  It is an extremely luminous
super star cluster, whose presence strongly suggests that the M82
starburst was quite active 60~Myr ago.  These clusters also offer
useful test paths through the ISM of M82, and in
addition to the well-known atomic absorption of Na\,I~D and Ca\,II
H$+$K, we find a probable DIB feature.

\section*{Acknowledgments}
We thank Claus Leitherer for providing model equivalent widths of
the Ca\,II triplet lines. We especially thank the staff of the Royal
Greenwich Observatory for obtaining the observations of M82 for us
through the service programme.
The William Herschel Telescope is operated on the island of La Palma
by the Royal Greenwich Observatory in the Spanish Observatorio del
Roque de los Muchachos of the Instituto de Astrof\'\i sica de Canarias.

\bsp
\label{lastpage}
\end{document}